\documentclass[3p,times]{elsarticle}
\usepackage{ecrc}

\volume{00}

\firstpage{1}

\journalname{Nuclear Physics A}

\runauth{Amir H. Rezaeian}

\jid{nupha}

\usepackage{amssymb}
\usepackage{graphicx}
\usepackage{amsmath}
\newcommand{\beq}{\begin{equation}}
\newcommand{\eeq}{\end{equation}}
\newcommand{\beqa}{\begin{eqnarray}}
\newcommand{\eeqa}{\end{eqnarray}}
\newcommand{\bseq}{\begin{subequations}}
\newcommand{\eseq}{\end{subequations}}

\def\x{{\boldsymbol x}}

\def\0{{\boldsymbol 0}}

\def\be{\begin{equation}}
\def\ee{\end{equation}}
\def\bea{\begin{eqnarray}}
\def\eea{\end{eqnarray}}

\def\fig#1{{Fig.~\ref{#1}}}

\def\simle{\mathrel{\rlap{\raise 0.511ex \hbox{$<$}}{\lower 0.511ex 
\hbox{$\sim$}}}}
\def\simge{\mathrel{ \rlap{\raise 0.511ex 
\hbox{$>$}}{\lower 0.511ex \hbox{$\sim$}}}}

\usepackage[figuresright]{rotating}

\begin{document}

\begin{frontmatter}


\title{Particle Production in pA Collisions and QCD Saturation}

\author{Amir H. Rezaeian}

\address{Departamento de F\'\i sica, Universidad T\'ecnica
Federico Santa Mar\'\i a, Avda. Espa\~na 1680,
Casilla 110-V, Valparaiso, Chile}

\begin{abstract}
The forthcoming LHC measurements  in proton-nucleus (pA) collisions at forward rapidities can discriminate between the color glass condensate (CGC) and alternative approaches including standard collinear factorization one.  We report some of our recent predictions based on gluon saturation/CGC formalism for pA collisions at the LHC including the charged hadron multiplicity distribution,  the nuclear modification factor for single inclusive hadron and prompt photon production, and the azimuthal angle correlation of the semi-inclusive hadron-photon production. 

\end{abstract}


\end{frontmatter}

\section{Introduction}
The CGC formalism is a self-consistent effective perturbative QCD theory at high energy (or small $x$) in which one
systematically re-sums quantum corrections which are enhanced by large
logarithms of $1/x$ and also incorporates non-linear high gluon
density effects which are important where the physics of gluon
saturation is dominant, for a pedagogical review, see Ref.\,\cite{CGC}. 
The greatest simplicity of the CGC formalism is the fact that the complexity of non-linear many-body problem at high energy is reduced into a one-scale problem, with a hard saturation scale $Q_s$ as the only dimensional relevant scale at which gluons recombination effects start to become as important as the gluon radiation. In the CGC approach the main features of particle production at high energy remain universal and are controlled by the saturation scale.  Although RHIC and the LHC forward rapidities both are within the kinematics interest of small-x region, the available kinematics phase space for particle production at the LHC is significantly larger than RHIC due to the larger energy of the collisions. Therefore, the LHC pA run can provide complementary information  to clarify the underlying dynamics of forward rapidity particle production at small $x$. 

The CGC formalism has been successfully applied to many processes in high energy collisions. Examples are structure functions (inclusive and diffractive) in Deeply Inelastic Scattering of electrons on protons or nuclei, and particle production in proton-proton, proton-nucleus and nucleus-nucleus collisions at RHIC, see Ref.\,\cite{CGC} and references therein. 
There are however alternative phenomenological approaches which combine nuclear shadowing, transverse momentum broadening and cold matter energy loss to describe the RHIC data at forward rapidities. The nuclear shadowing, saturation and transverse momentum broadening have the same origin and are consistently incorporated in the CGC formalism from a first principle approach. However, the energy loss effect, medium modified fragmentation processes and in general, possible initial state-final state interference effects in pA collisions have not been yet incorporated in the CGC formalism (in this sense, the CGC results only contain the genuine initial-state effects).   

The CGC formalism has been also successful in describing the first LHC data in proton-proton (pp) and nucleus-nucleus (AA) collisions \cite{m1,j1,m2,m-mul,IP,KLN}. 
Here we show some of our recent quantitative predictions for the LHC pA  collisions \cite{m-mul,m-h,m-p} which can discriminate between the CGC and alternative approaches. 
The details of results shown here can be found in Refs.\,\cite{m1,m-mul,m-h,m-p}.

\begin{figure}[t]     
\begin{center}      
                               \includegraphics[width=5.5 cm] {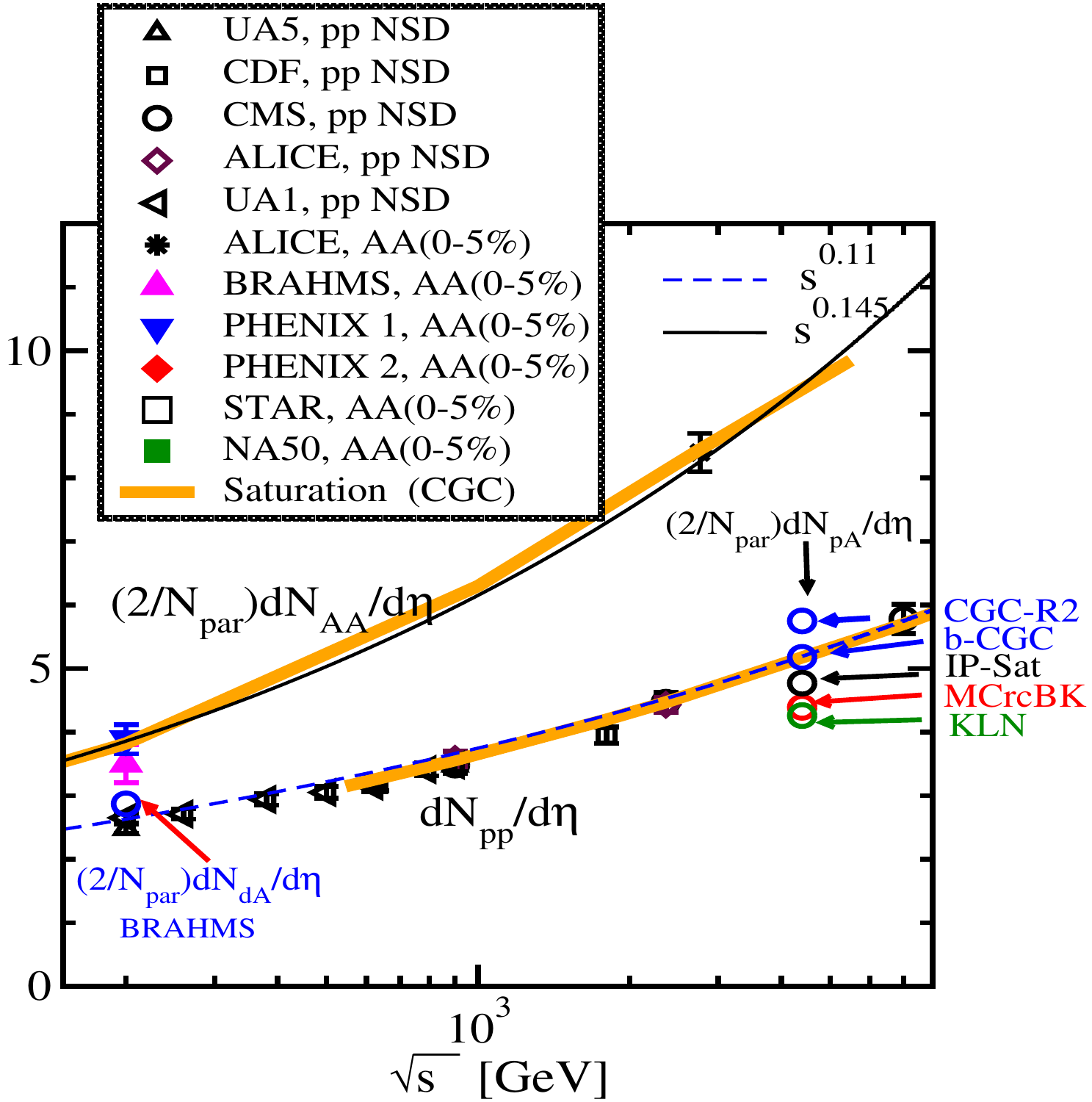}              
                                 \includegraphics[width=5.5 cm] {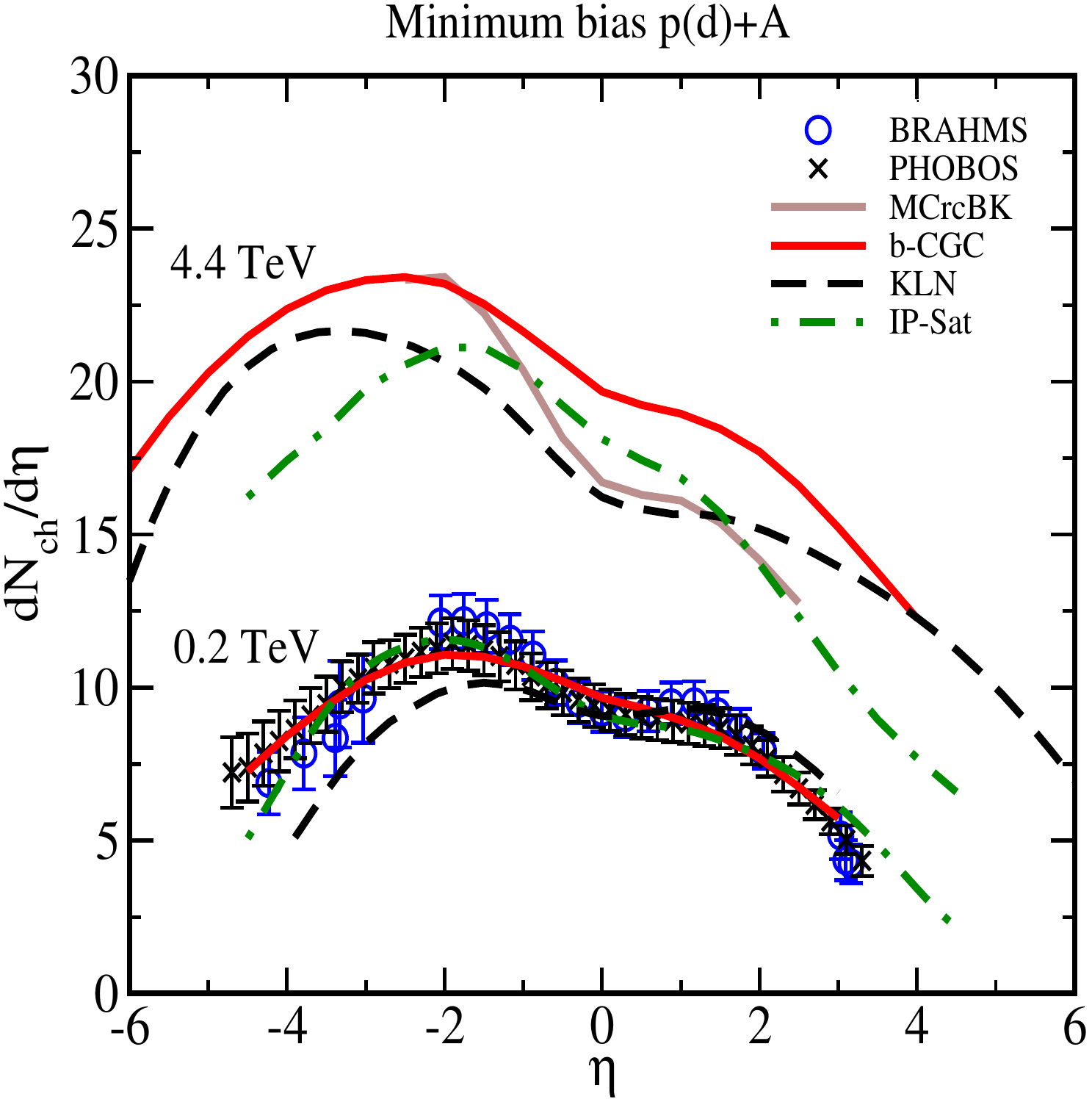} 
\caption{Right: Pseudo-rapidity distribution in p(d)A collisions calculated in the CGC formalism with different saturation models.  Left: the corresponding energy-dependence of the scaled multiplicity of the charged particles production in minimum bias pA (and pp and AA \cite{m1}) collisions. }
\label{fig-1}
\end{center}
\end{figure}

\section{Some predictions}
In \fig{fig-1}(right), we show various predictions for pseudo-rapidity distribution in p(d)A collisions at the LHC based on the CGC formalism. The theoretical curves labeled by MCrcBK \cite{j1}, b-CGC \cite{m-mul}, IP-Sat \cite{IP} and KLN \cite{KLN}  are all based on the leading log $k_t$-factorization formalism \cite{kt} but different saturation models. These saturation models are most popular ones. Theoretical uncertainties about $10-20\%$ are implicit in all curves shown in \fig{fig-1}(right).  One of the intriguing feature of the recent LHC data has been
the very different power-law energy behavior of charged hadron
multiplicities in AA compared to pp collisions \cite{m1}.  In \fig{fig-1}(left), we show the corresponding energy-dependence of the multiplicity scaled with number of participant $N_{par}$ in pA collisions coming from different saturation models (we take $N_{par}=7.6$  for minimum bias pA collisions at $\sqrt{s}=4.4$ TeV). As a comparison, we also show in the same plot the corresponding curve calculated in Ref.\,\cite{m1} for pp and AA collisions\footnote{The theoretical data point labeled CGC-R2 is based on a prediction  which incorporates the MLLA gluon decay cascade effect into the leading log $k_t$-factorization \cite{m-mul}.}.  Within theoretical uncertainties the above mentioned saturation models provide rather consistent predictions for pA collisions at the LHC. Nevertheless, high precision LHC pA data on the charged particles pseudo-rapidity multiplicity distribution can in principle discriminate between these models and also examine the $k_t$-factorization at an unprecedented level. Note that the $k_t$-factorization was only proven for dilute-dense scatterings such as high-energy pA collisions \cite{kt}, see also Ref.\,\cite{kt-2}.
\begin{figure}[t]  
\begin{center}     
                                \includegraphics[width=5.5 cm] {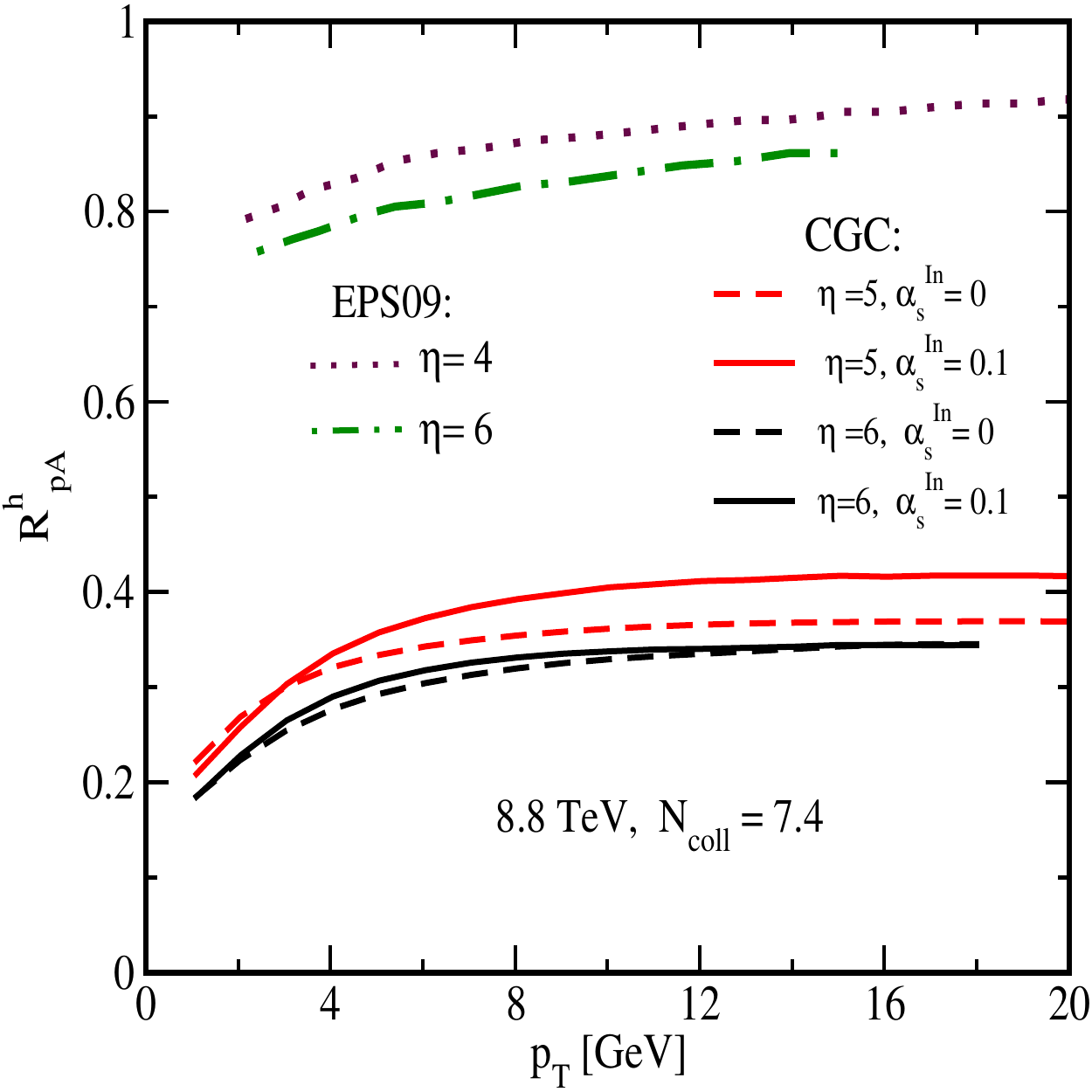} 
                                  \includegraphics[width=5.5 cm] {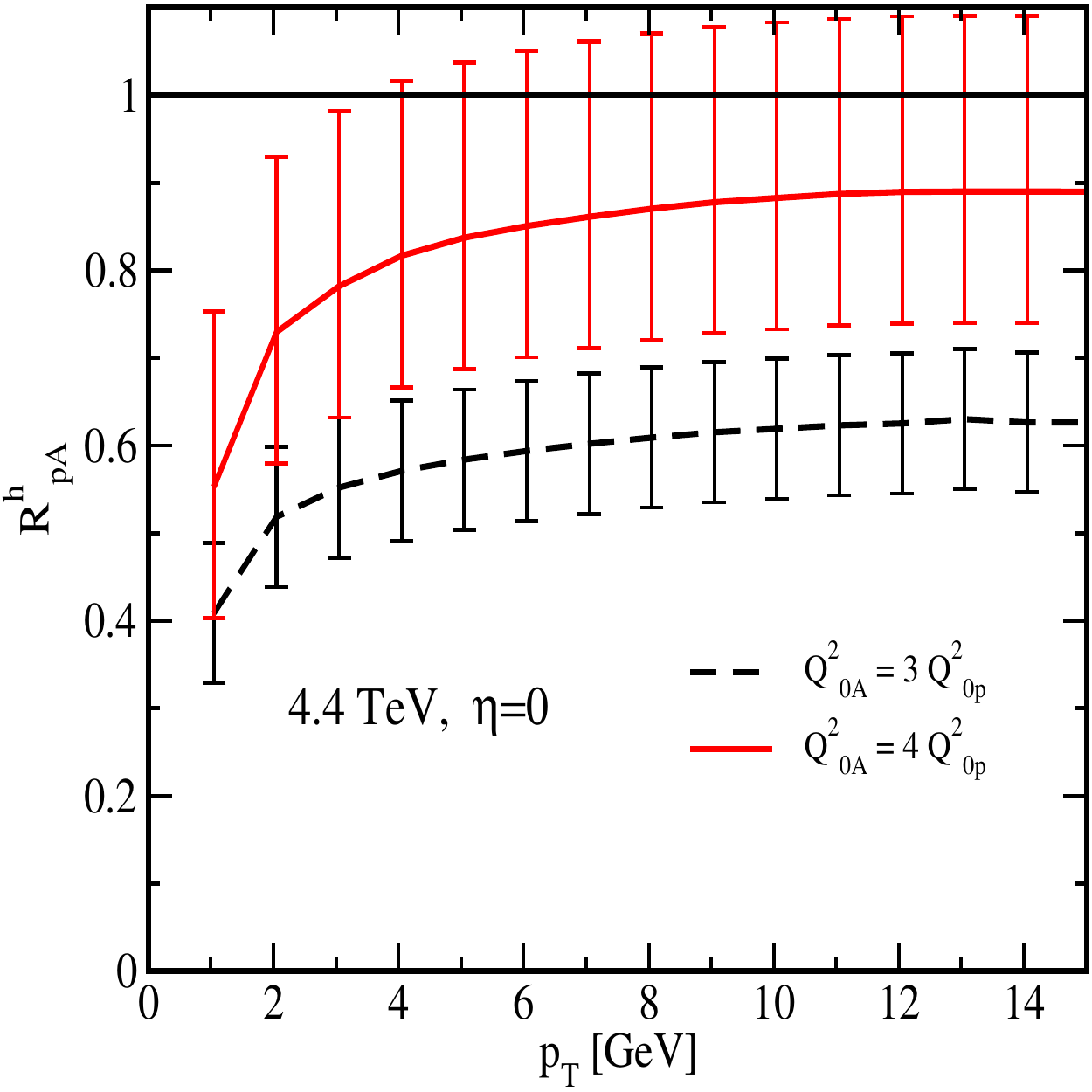}
\caption{ $R^h_{pA}$ for $h^{+}+h^{-}$ production in pA collisions at the LHC at $\eta=0$ (right) and forward rapidities (left) obtained with different initial saturation scale $Q_{0A}$ and the strong coupling constant $\alpha^{in}_s$ in the inelastic terms \cite{m-h}. The curves labeled EPS09 are based on the collinear factorization \cite{pqcd-h}. }
\label{fig-lhc4}
\end{center}
\end{figure}
\begin{figure}[!h]
\begin{center}
                    \includegraphics[width=5.5 cm] {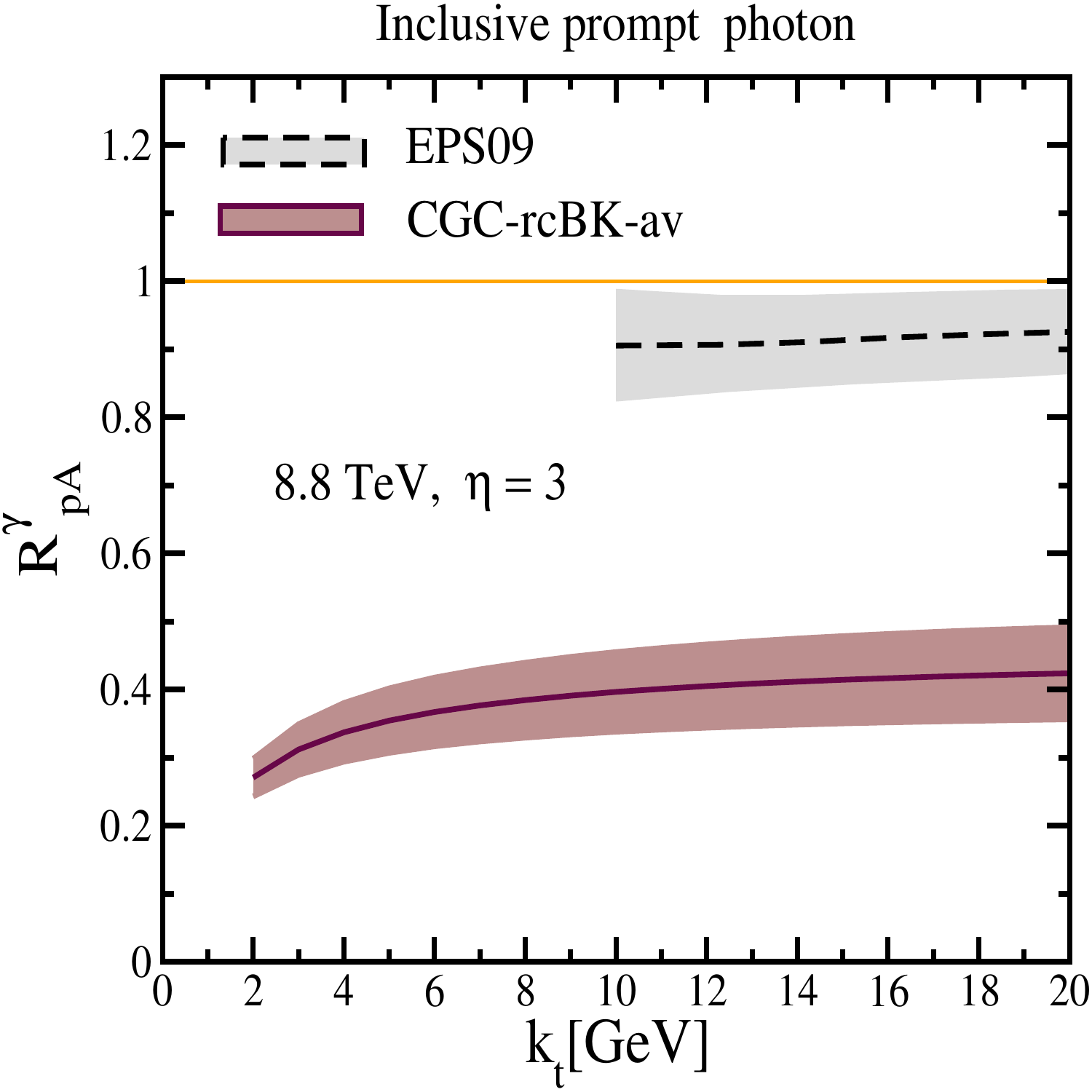}                            
                                  \includegraphics[width=5.5 cm] {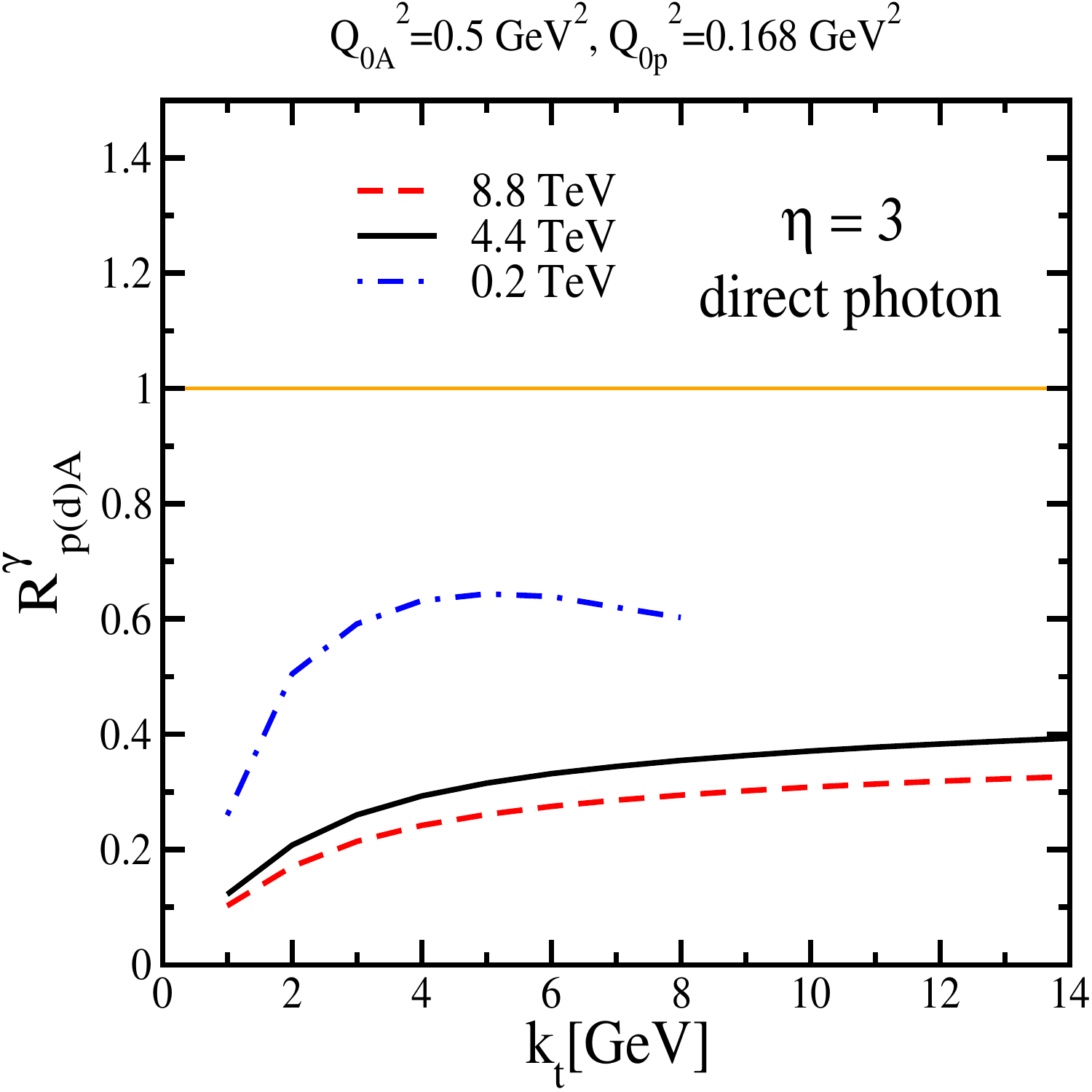}        
\caption{Right: nuclear modification factor for direct photon production at $\eta=3$ in minimum-bias p(d)A collisions at RHIC and the LHC. Left: the inclusive prompt photon $R^{\gamma}_{pA}$ from the CGC (CGC-rcBK-av) \cite{m-p} and the collinear factorization results (EPS09) \cite{pqcd-p} at $\eta=3$ at the LHC. }
\label{fig-r3}
\end{center}
\end{figure}

In nuclear collisions, nuclear effects on single particle production are usually evaluated in terms of ratios of particle yields in pA and pp collisions scaled with number of binary collisions $N_{coll}$, the so-called  nuclear modification factor $R^{h}_{pA}$ for hadron ($R^{\gamma}_{pA}$ for photon). We take $N_{coll}=3.6, 6.5$ and $7.4$ at $\sqrt{s}=0.2, 4.4$ and $8.8$ TeV, respectively. 
In  \fig{fig-lhc4},  we show our predictions \cite{m-h} for $R^h_{pA}$ for inclusive charged
hadron production at $\sqrt{s}=4.4$ and $8.8$ TeV at midrapidity (right) and forward rapidities (left)
obtained from the running-coupling Balitsky-Kovchegov (rcBK) evolution equation \cite{rcbk} assuming two different initial nuclear
saturation scales of $Q_{0A}^2 = 0.5, 0.67\, \text{GeV}^2$ and a fixed initial saturation scale for proton 
$Q_{0p}^2 = 0.168~\text{GeV}^2$.  The assumed initial saturation scales $Q_{0p}$ and $Q_{0A}$ are extracted from a fit to small-x experimental data and consistent with the existing data from HERA, RHIC and the LHC at small-x \cite{j1,m-h,m-p}. 
The theoretical error bars in \fig{fig-lhc4} show the uncertainties mainly associated with
the choice of the strong-coupling $\alpha_s^{\text{In}}$ in the inelastic contribution \cite{m-h,alex}. We note that there
are large uncertainties in $R^h_{pA}$ at midrapidity at the LHC due to
the choice of the initial saturation scale for the rcBK evolution
equation, and the value of strong-coupling constant in the higher order inelastic terms \cite{m-h,alex}. More importantly, large sensitivity of $R^h_{pA}$ to the
value of $\alpha_s^{\text{In}}$  at midrapidity at the LHC indicates
that higher order corrections should be important at midrapidity at the LHC
energy \cite{m-h,alex}. Therefore, we believe that  the current CGC predictions for
$R^h_{pA}$ at midrapidity may be less reliable compared to the 
results for the very forward rapidity collisions. In \fig{fig-lhc4} (left), we also compare the CGC predictions for $R^h_{pA}$ at $\sqrt{s}=8.8$ TeV at forward rapidities  with the collinear factorization results (EPS09) \cite{pqcd-h}.

There are advantages to studying prompt photon production as compared to hadron production. It is theoretically cleaner; one avoids the difficulties involved with description of hadronization.  Also, one does not have to worry about possible initial state-final state interference effects which may be present for hadron production \cite{m-p,m-p-m}.  In \fig{fig-r3} (right), we compare $R_{p(d)A}^{\gamma}$ for direct photon at $\eta=3$ at RHIC and the LHC. The large suppression of $R_{pA}^{\gamma}$ at the LHC is impressive given the fact that a good amount of the suppression of $R^{\gamma}_{dA}$ at RHIC is due to the projectile being a deuteron rather than a proton (isospin effect) \cite{m-p}. In \fig{fig-r3} (left), we compare the CGC prediction (CGC-rcBK-av)  \cite{m-h} with the collinear factorization result (EPS09) \cite{pqcd-p} for inclusive prompt photon $R_{pA}^{\gamma}$  at $\eta=3$ at the LHC.

From Figs.\,(\ref{fig-lhc4},\ref{fig-r3}), it is seen that the suppression of the nuclear modification factor at the LHC forward rapidities for both hadron and prompt photon, is larger in the CGC \cite{m-h,m-p} compared to  the collinear factorization (standard parton model)  approach \cite{pqcd-h,pqcd-h} (see also Ref.\,\cite{pqcd-o}).  Therefore, the LHC measurements of the single inclusive hadron and prompt photon at very forward rapidities can clearly discriminate between the collinear and the CGC approach and provide direct evidence in favor of importance of the gluon saturation and small-x resummation. 

\begin{figure}[t]      
\begin{center} 
                               \includegraphics[width=5.5 cm] {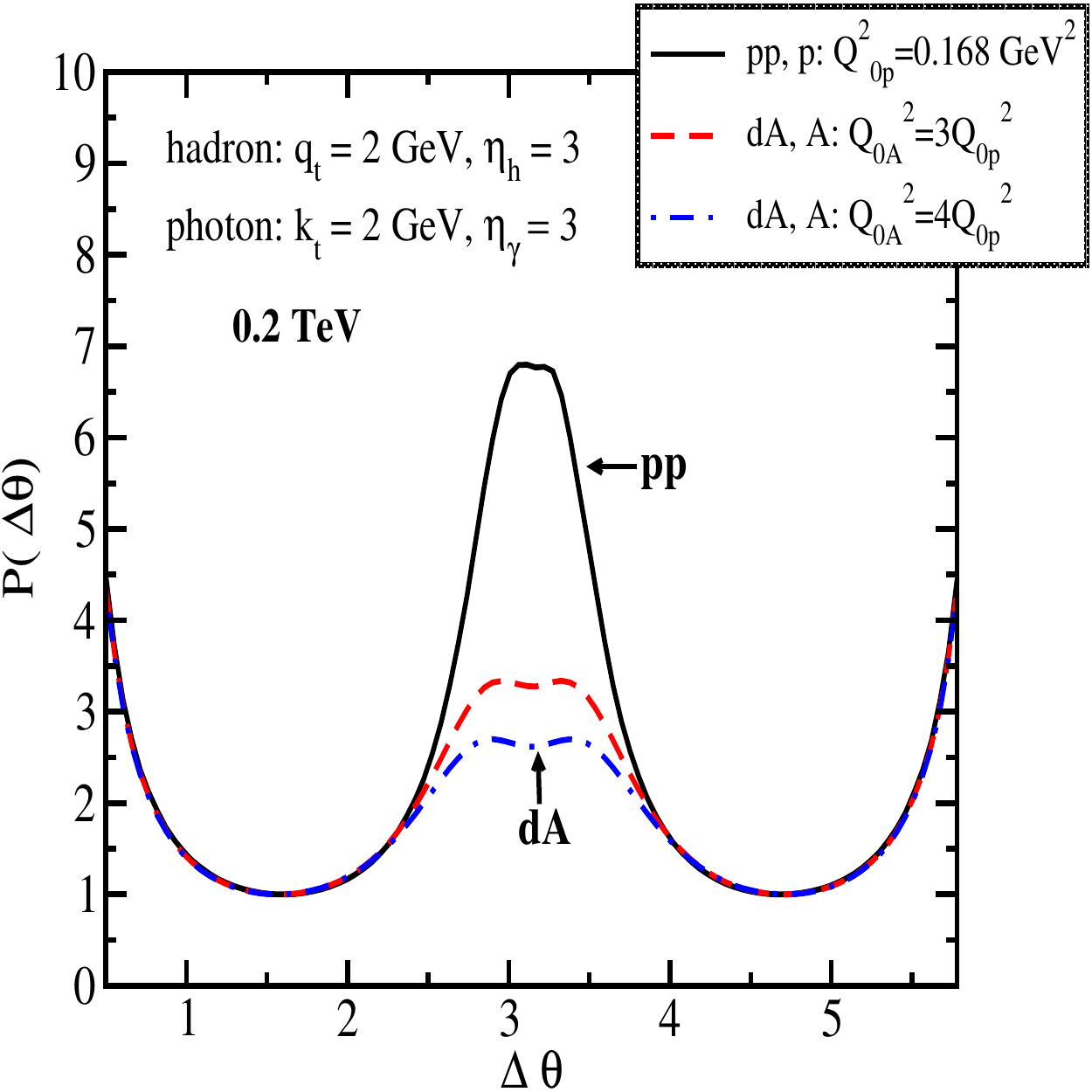}                 
                                  \includegraphics[width=5.5 cm] {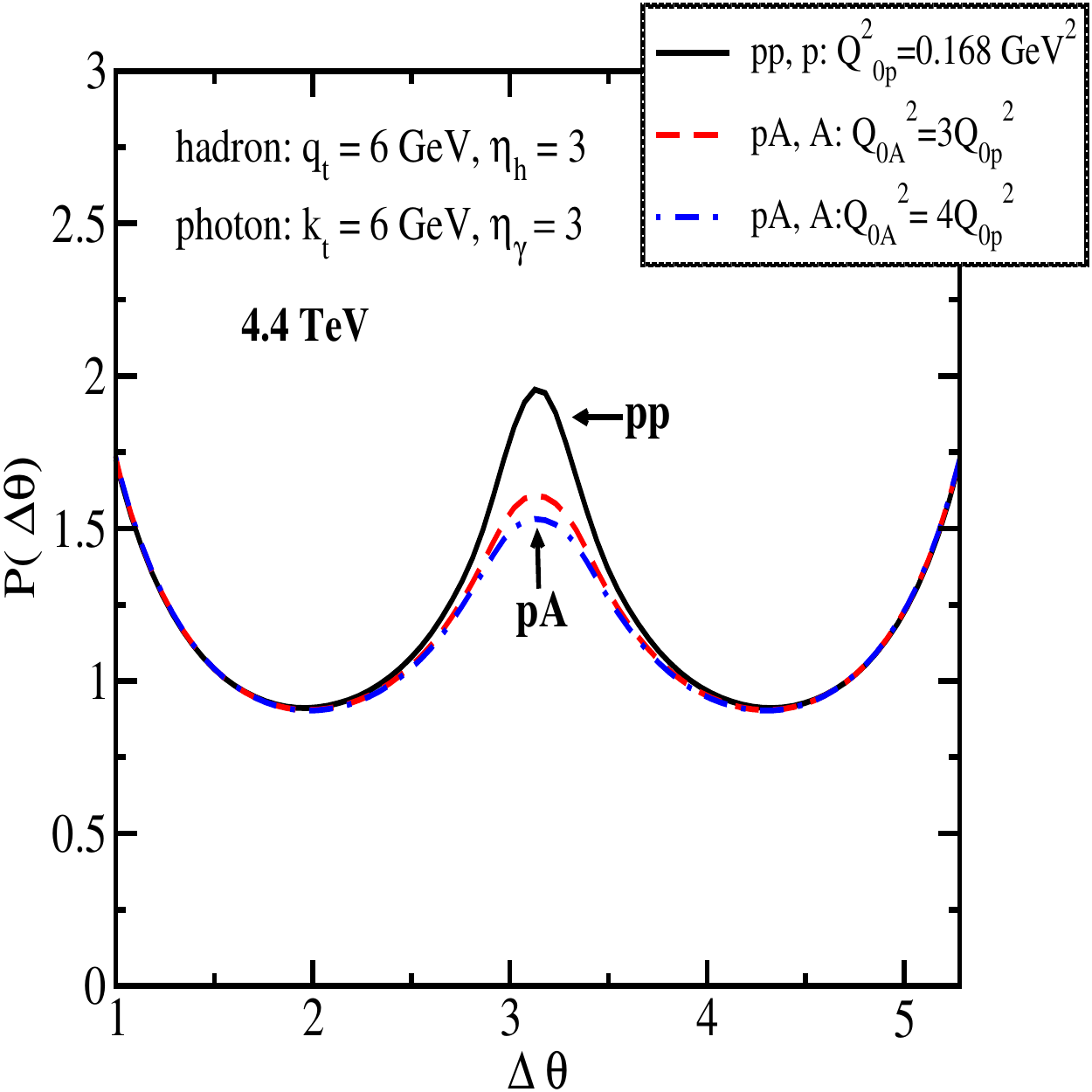}  
\caption{The relative azimuthal correlation $P(\Delta \theta)$ for minimum-bias p(d)A and pp collisions at RHIC and the LHC .}
\label{fig1-1}
\end{center}
\end{figure}

Finally, in \fig{fig1-1}, we show azimuthal angle correlations of the prompt photon-hadron  production, where the 
angle $\Delta \theta$ is the difference between the azimuthal angle of the measured hadron and single prompt photon. The correlation $P(\Delta \theta)$ is defined as the probability of, the single semi-inclusive prompt photon-hadron production at a certain kinematics and angle  $\Delta \theta$ given the  production with the same kinematics at  a fixed reference angle $\Delta \theta_c=\pi/2$ \cite{m-p}. Note that $P(\Delta \theta)$ defined in this way is different from  the so-called coincidence probability\footnote{A detailed paper to analyze the coincidence probability for the semi-inclusive photon-hadron production in pA collisions is under preparation.}. 
It is clear that the away-side prompt photon-hadron cross-section (at $\Delta \theta\approx \pi$)  is suppressed for a bigger saturation scale (corresponding to a denser system).   The suppression of the away-side azimuthal photon-hadron correlations  with decreasing the transverse momentum of the produced prompt photon or hadron, or increasing the energy, or increasing the size/density of system, all uniquely can be explained within the universal picture of gluon saturation without invoking any new parameters or ingredients to our model \cite{m-p}.  We emphasize that prompt photon-hadron azimuthal angular correlations suffers from much less theoretical uncertainties as compared to di-hadron azimuthal angular correlations which involve higher number of Wilson lines. 

\vspace{0.2cm}
\noindent{\bf Acknowledgements\ \ }
The author would like to thank Jamal Jalilian-Marian and Genya Levin for the collaboration on some of topics reported here. 
This work is supported in part by Fondecyt grants 1110781.










\end{document}